\documentclass[%
mathleft,%
final,%
]{an}
\usepackage{graphicx}
\usepackage{times}
\newcommand{\kms}{km\,s$^{-1}$}

\begin{document}

\def\ha{H$\alpha$}
\def\kms{km~s$^{-1}$}
\def\msun{M$_\odot$}
\def\rsun{R$_\odot$}
\def\lsun{L$_\odot$}
\def\mjup{M$_{\rm Jup}$}
\def\teff{$T_{\rm eff}$}
\def\logg{$\log g$}
\def\lbol{$L_{\rm bol}$}
\def\lha{$L_{{\rm H}\alpha}$}
\def\lx{$L_X$}

\def\2m{2M0535$-$05}

% The following seven commands are intended for editorial usage and
% should be ignored by the author(s).
\Pagespan{1}{}% Document's page range.
% If second parameter is left empty, the last page is computed
% automatically.
\Yearpublication{2013}%
\Yearsubmission{2012}%
\Month{1}%
\Volume{334}%
\Issue{1}%
\DOI{This.is/not.aDOI}%

\title{An Empirical Correction for Activity Effects on the Temperatures, 
Radii, and Estimated Masses of Low-Mass Stars and Brown Dwarfs}

\author{Keivan G.\ Stassun\inst{1,2}\fnmsep\thanks{Corresponding author. 
  \email{keivan.stassun@vanderbilt.edu}},
% Example for footnote, note the usage of the \fnmsep command
% as separator between institute number and footnote mark}
Kaitlin M.\ Kratter\inst{3},
Aleks Scholz\inst{4},
Trent J.\ Dupuy\inst{3}
}
\titlerunning{Correction for Activity Effects on Stellar Properties}
\authorrunning{Stassun et al.}
\institute{
\inst{1}Department of Physics \& Astronomy, Vanderbilt University,
VU Station B 1807, Nashville, TN USA \\
\inst{2}Fisk University \\
\inst{3}Harvard-Smithsonian Center for Astrophysics \\
\inst{4}Dublin Institute for Advanced Studies}

\received{XXXX}
\accepted{XXXX}
\publonline{XXXX}


\abstract{
We present empirical relations for 
%\sout{correcting the estimated} \sout{quantitatively} 
determining the amount by which the 
effective temperatures and radii---and therefore the estimated masses---of low-mass stars and brown dwarfs
are altered due to chromospheric activity. 
%A significant fraction of M-dwarfs show evidence of chromospheric activity. Their measured physical properties are
%known to be discrepant with both their inactive counterparts, and theoretical models. 
%Here we derive an empirical relation that links both 
Our relations are based
on a large set of well studied low-mass stars in the field 
%with \ha\ activity measurements, 
%and known distances, 
and on 
%We complement this sample with 
a set of benchmark low-mass eclipsing binaries.
%with X-ray activity measurements
%and with directly measured temperatures and radii. 
%For these objects, we link the temperature suppression and radius inflation
%to the strength of X-ray emission and then 
%from which we indirectly infer the
%\ha\ activity. 
%Both samples yield consistent relations
The relations link the amount by which an active object's temperature is suppressed, 
and its radius inflated, to the strength of its \ha\ emission. 
%\ha\ emission to temperature suppression and radius inflation.
%\sout{Bolometric luminosity is found to be approximately preserved by these temperature and radius corrections.}
These relations are found to approximately preserve bolometric luminosity.
We apply these relations to the peculiar brown-dwarf
eclipsing binary \2m, in which the active, higher-mass brown dwarf has a
cooler temperature than its inactive, lower-mass companion. %\sout{We find that }
%\sout{the \ha-corrected
%temperatures bring the inferred masses of the brown dwarfs into agreement with theoretical isochrones}
The relations correctly reproduce the observed temperatures and radii of \2m\ after accounting for the \ha\ emission;
%\sout{our relations precisely explain the observed temperatures and radii of \2m\ from the observed \ha\ emission;}
\2m\ would be in precise agreement with theoretical isochrones were it inactive.
%\sout{Our empirical relations}
The relations that we present are applicable to brown dwarfs and low-mass stars with masses below 0.8 \msun\
and for which the activity, as measured by \ha, is in the range
$-4.6 < \log$~\lha/\lbol\ $< -3.3$. 
We expect these relations to be most useful for correcting
%\sout{temperatures and} 
radius and mass estimates of low-mass stars and brown dwarfs over their active 
lifetimes (few Gyr). 
%\sout{and thereby also the inferred masses of objects with unknown} 
We also discuss the implications of this work for determinations of young cluster IMFs.
%and when the ages or distances (and therefore luminosities) are unknown. 
}

\maketitle

\section{Introduction\label{intro}}
Observational evidence strongly indicates that the fundamental properties of low-mass stars 
can be altered in the presence of strong magnetic activity 
%that is common to low-mass objects
(e.g., Morales et al.\ 2008; L\'opez-Morales 2007; Ribas 2006).
In particular, observations of active, low-mass eclipsing binary (EB)
stars have found the empirically measured stellar radii ($R$) to be inflated by $\approx$10\%, 
and the empirically measured stellar effective temperatures
(\teff) to be suppressed by $\approx$5\%, relative to the predictions of
standard theoretical stellar evolution models, 
which better match the properties of inactive objects
(see Coughlin et al.\ 2011; Kraus et al.\ 2011; Morales et al.\ 2010; 
Stassun et al.\ 2009, and references therein).
%Because the mass-radius and mass-\teff\ relationships are central to
%our understanding of stellar evolution, 
Resolving these discrepancies
will be critical to the ongoing development of accurate theoretical 
stellar models (see Stassun et al.\ 2010). 
Accurate estimates of stellar radii are especially
important in the context of searches for transiting exoplanets,
which rely upon the assumed stellar radius/density to infer the planet radius/density.

\begin{figure}[ht]
%\epsscale{0.75}
\includegraphics[angle=90,width=\linewidth,trim=0mm 60mm 15mm 5mm,clip]{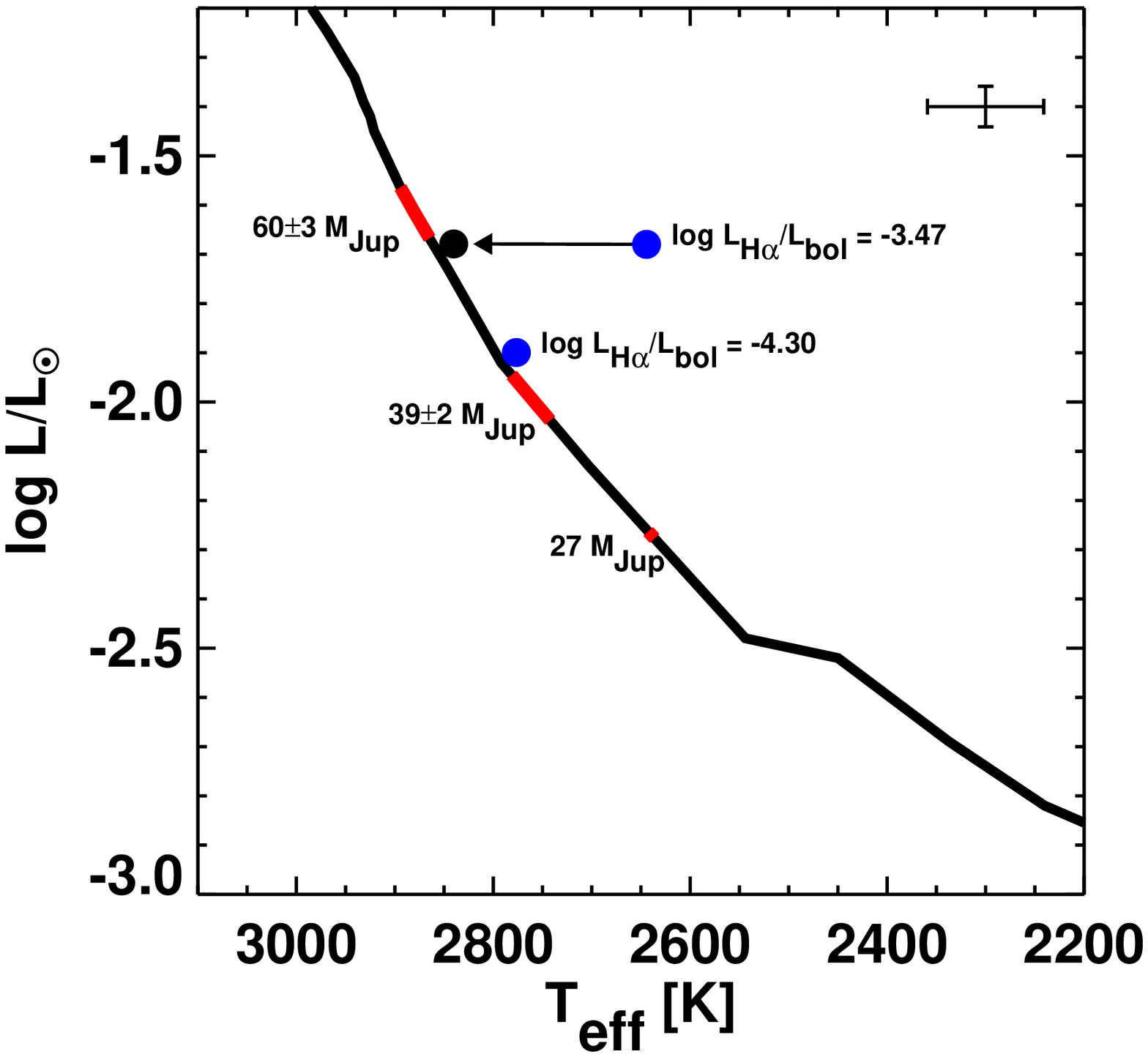}
\caption{\label{fig:2m0535} 
H-R diagram for the primary and secondary BD components of the 
EB \2m\ in the $\sim$1-Myr Orion Nebula Cluster (Stassun et al.\ 2006, 2007).
The measured \teff\ and \lbol\ 
%(the latter calculated from the directly measured \teff's and radii) 
for both BDs are represented as blue symbols. 
Measurement uncertainties in \teff\ and \lbol\ are represented by the error bars at upper right.
The dynamically measured masses of the primary and secondary 
%are 60$\pm$3 and 39$\pm$2 \mjup, respectively, 
are represented as red bars on the 1-Myr theoretical isochrone of Baraffe et al.\ (1998).
The measured \lha/\lbol\ for the two components are indicated next to the
blue symbols. 
%The inactive secondary appears close to its expected position on the isochrone, whereas 
The active primary appears far cooler than expected, and therefore
appears to be much younger than the secondary and to have a mass of only $\approx$27 \mjup\ 
based on its observed \teff, a factor of 2 lower than its true mass. 
Shifting the position of the active primary (arrow) using our empirically calibrated \ha-based
relations 
%for \teff\ suppression and radius inflation 
brings the primary into much closer
agreement with its theoretically expected position in the HR diagram (black symbol);
this is where the active primary would be if it were not active.
}
\end{figure}

Activity effects also lead to errors in object masses ($M$) when these are derived from \teff.
A particularly salient example is \2m, an EB
in the Orion Nebula Cluster (age $\sim$1 Myr)
comprising two brown dwarfs (Stassun et al.\ 2006, 2007).
The primary and secondary brown dwarf (BD) components of \2m\ have dynamically measured 
masses of 60$\pm$3 and 39$\pm$2 \mjup, respectively, 
and \teff\ ratio of $T_1 / T_2 = 0.952 \pm 0.004$ (G\'omez Maqueo Chew et al.\ 2009).
That is, the 
system exhibits a reversal of the usual $M$-\teff\ relation, such that
the primary component is cooler than its companion.
%This behavior is not predicted by theoretical models for coeval BDs. 
Figure~\ref{fig:2m0535} shows the \2m\ system on the Hertzsprung-Russell (H-R) diagram. 
%compared to the 1~Myr isochrone of Baraffe et al.\ (1998).
The secondary BD's \teff\ and bolometric luminosity (\lbol, calculated directly
from the empirically measured \teff\ and $R$) place it at a position
that is consistent with that predicted by the model isochrone.
In contrast, the primary is far displaced from its expected position, and so
{\it appears} to have a mass of only $\sim$25 \mjup---more than a factor of 2 
lower than its true mass---on the basis of its low \teff.
Reiners et al.\ (2007) used spectrally resolved H$\alpha$ measurements to show 
that, whereas the secondary BD in \2m\ is chromospherically quiet, the 
primary BD is highly chromospherically active, perhaps a consequence of its
rapid rotation. 
%(G\'omez Maqueo Chew et al.\ 2009). 
%Thus, magnetic activity in the primary BD
%could be responsible for its highly suppressed \teff, similar to what has 
%been seen for low-mass stellar EBs in the field. 

%Given the exhibition of activity-related \teff\ suppression in \2m, the 
%first and only EB containing an active BD, it is 
%likely that this phenomenon extends to other active BDs and low-mass stars in star 
%forming regions and in the field.  
Since magnetic activity seems to alter the fundamental properties of both stars and BDs,
it would be valuable to have an easily observable metric with which to 
quantitatively assess the degree to which a given object's \teff\ has been suppressed and its
radius inflated. 
Here we present such an empirical metric by relating the degree of \teff\ 
suppression and radius inflation to the strength of \ha\ emission, 
a commonly used and readily 
observable tracer of chromospheric activity (Scholz et al.\ 2007; Berger 2006).

\section{Methods and Data Used\label{methods}}

We use two different samples and approaches 
to empirically determine a relationship between \teff\ suppression, $R$
inflation, and the level of activity as measured from \ha. 

\begin{figure}[ht]
%\epsscale{0.5}
\includegraphics[angle=90,width=\linewidth,trim=28mm 10mm 14mm 10mm,clip]{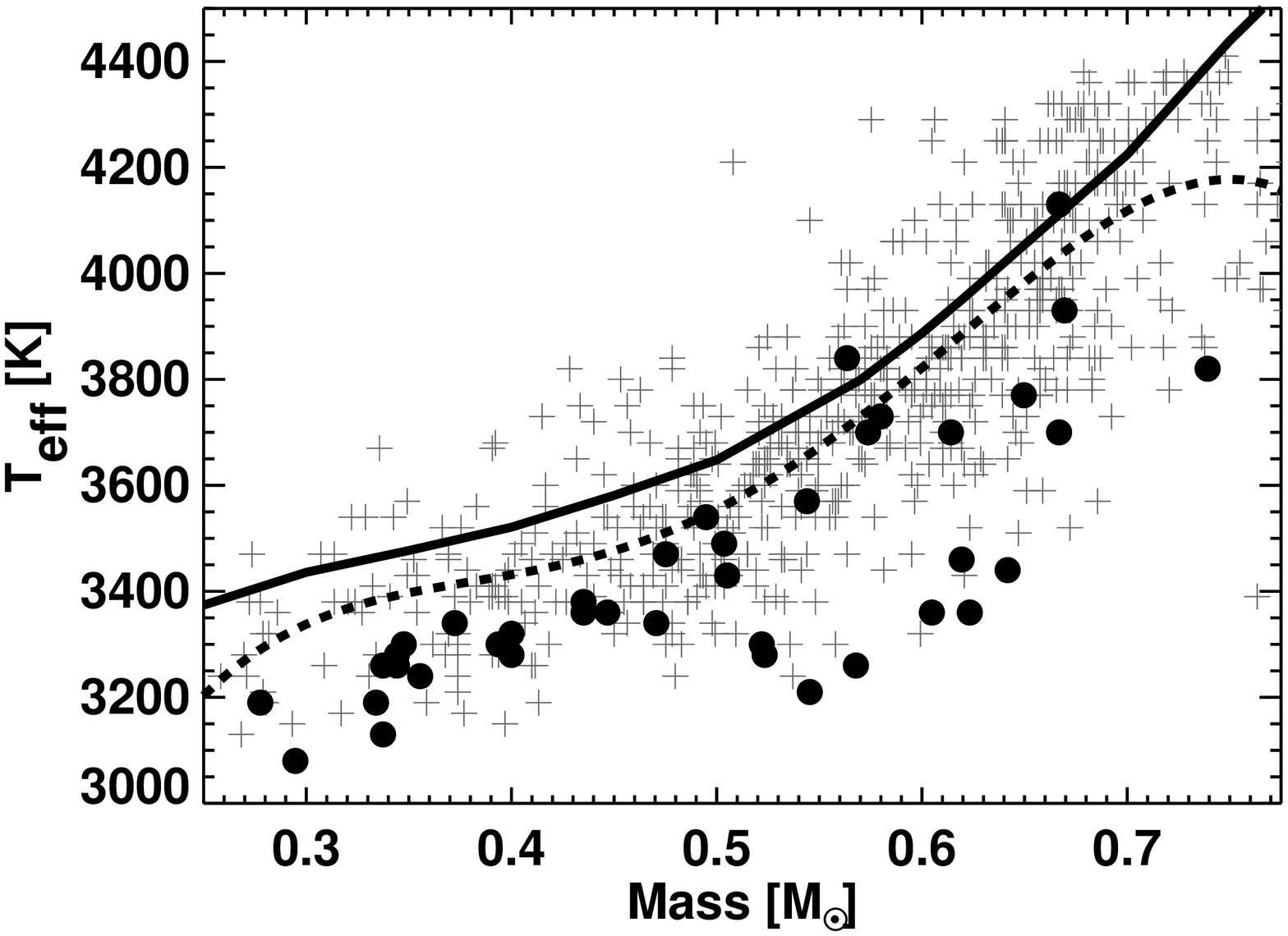}
\includegraphics[angle=90,width=\linewidth,trim=4mm 10mm 10mm 10mm,clip]{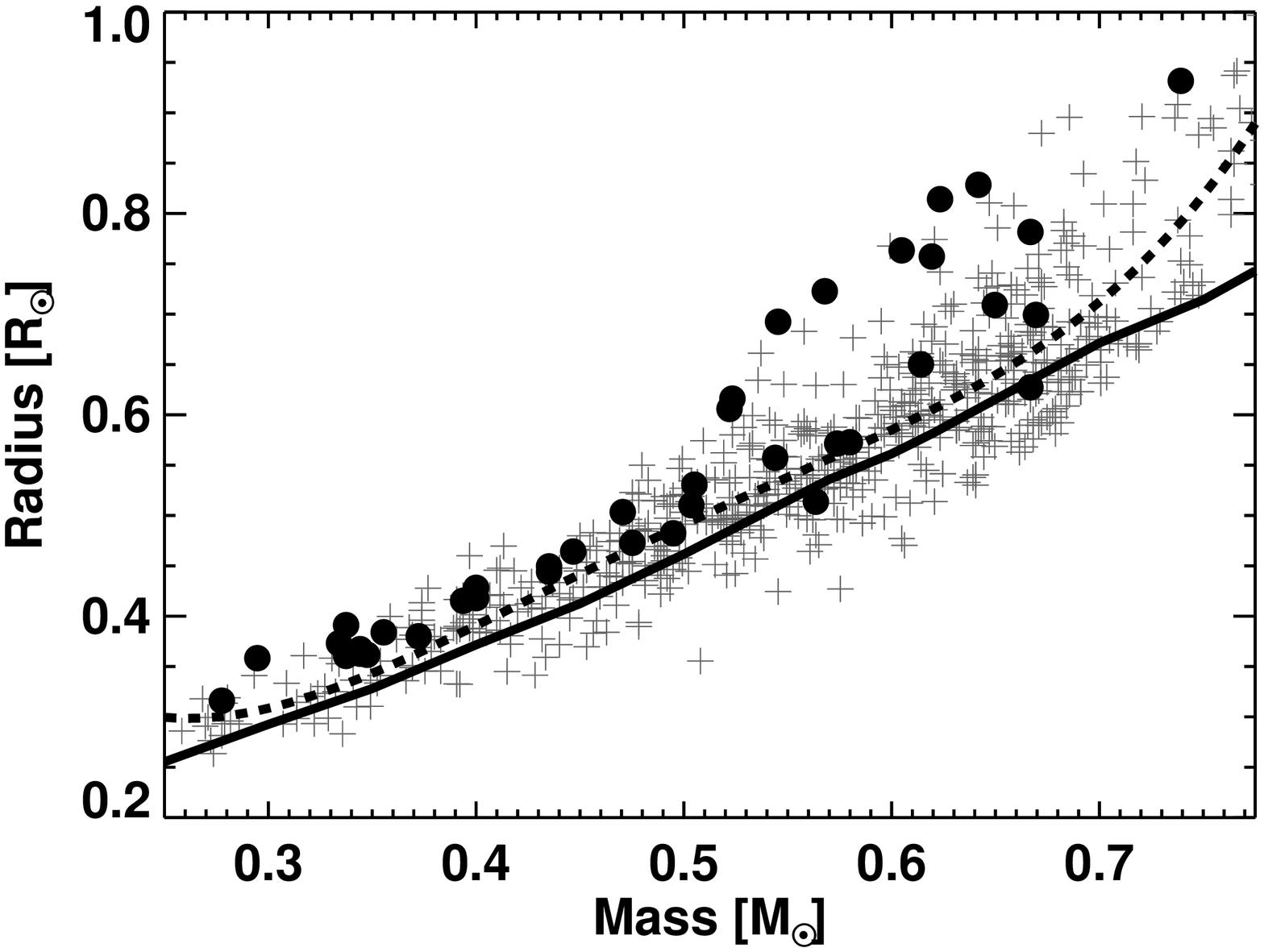}
\caption{\label{fig:pmsudata} 
\teff\ vs.\ Mass (top) and Radius vs.\ Mass (bottom) for M-dwarfs with 
trigonometric distances and \ha\ measurements from the PMSU catalog. 
Active objects (\ha\ in emission) are filled symbols.
Solid curve is a 3 Gyr isochrone (Baraffe et al.\ 1998).
Dashed curve is a polynomial fit to the non-active objects.
\ha-active dwarfs are significantly displaced to lower
\teff\ and larger radii compared to both the isochrone and to the non-active dwarfs.
}
\end{figure}

\begin{figure}[ht]
%\epsscale{0.5}
\includegraphics[angle=90,width=\linewidth,trim=28mm 5mm 14mm 15mm,clip]{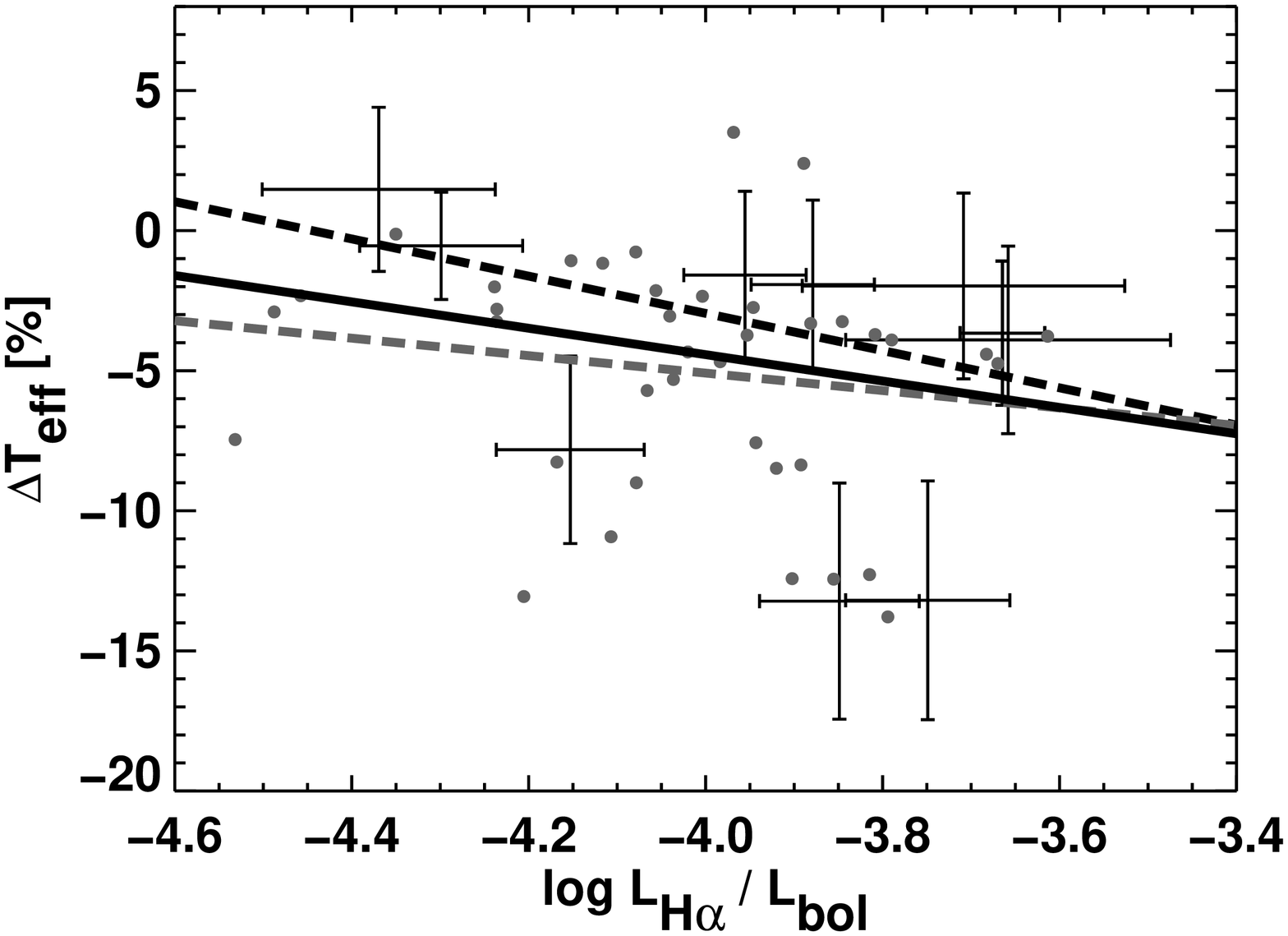}
\includegraphics[angle=90,width=\linewidth,trim=4mm 5mm 10mm 15mm,clip]{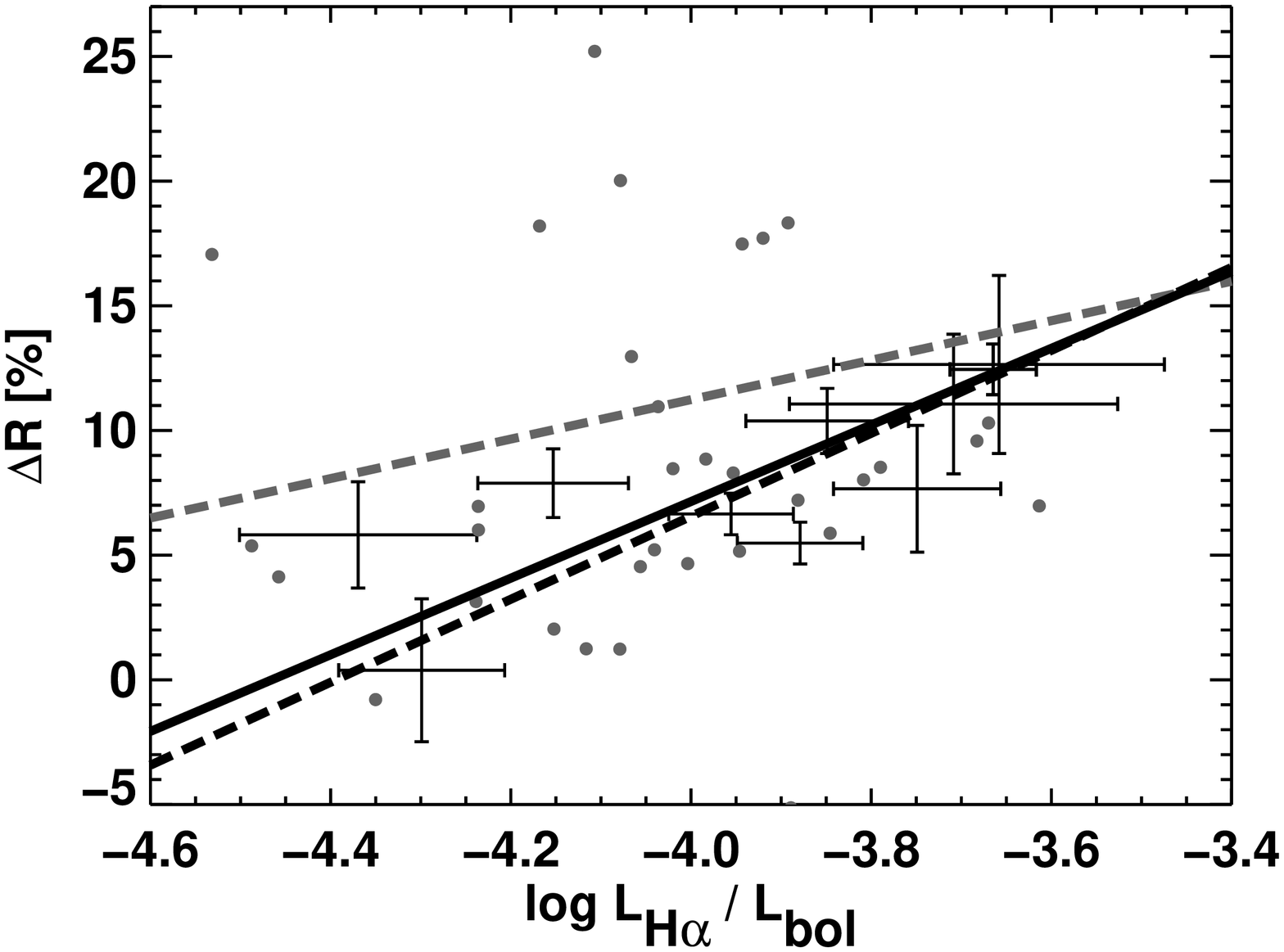}
\caption{\label{fig:finalrelations} 
\teff\ suppression (top) and radius inflation (bottom) as a function of fractional \ha\
luminosity for both field M-dwarfs (filled symbols) and low-mass eclipsing binaries (error bars). 
Linear fit relations to both sets are
dashed lines. The final averaged best-fit relation
the solid line.
See text for the linear fit coefficients.
%(Eqs.~\ref{eq:Trelation} and \ref{eq:Rrelation}).
}
\end{figure}

First 
%a large sample 
%of stars without direct mass, radius, or \teff\ measurements but 
%with direct distances, spectral types, and reliable \ha\ measurements.  
we use the large set of nearby field M dwarfs with well measured spectral 
types and \ha\ equivalent widths (EWs) from the PMSU
catalog (Reid et al.\ 1995; Hawley et al.\ 1996). 
Following Morales et al.\ (2008), we restrict 
ourselves to the sample of 746 stars with distances determined directly from 
trigonometric parallaxes.
Fig.~\ref{fig:pmsudata} shows the estimated \teff\ and $R$ of the PMSU sample stars
as a function of $M$.
Strongly chromospherically active stars---defined as those with \ha\ in emission---show
a clear displacement to lower \teff\ and to larger $R$ relative to both the 
theoretical isochrone
and to the non-active stars, whereas non-active stars more closely track the isochrone.
%The \ha\ active stars also show a displacement to larger $R$ relative to both the 
%theoretical isochrone and to the non-active stars. 
%The mean offset of the \ha\ active stars relative to the isochrone (solid curve) 
%is 10.0$\sigma$ for $R$ and $-11.1\sigma$ for \teff, where
%$\sigma$ is the standard deviation of the mean (i.e., r.m.s./$\sqrt{N}$). 
%The mean offset relative to the non-active stars (dashed curve) 
%is 6.8$\sigma$ for $R$ and $-7.5\sigma$ for \teff. 
From these \teff\ and $R$ offsets, and the observed \ha\ emission, we derive 
linear relationships between $\Delta$\teff, $\Delta R$, and $\log$~\lha/\lbol\
(see Fig.~\ref{fig:finalrelations}). The relationships are statistically 
significant with $>$90\% confidence.

%\subsection{Low-mass Eclipsing Binaries with X-ray Emission}

The second approach uses the much smaller sample of stars in EBs that have directly measured 
masses, radii, and reliable \teff, but for which we must use X-ray flux as a proxy for \ha.
We use the small set of low-mass EBs with accurately
measured $M$, $R$, \teff, and X-ray luminosities (\lx) from 
L\'opez-Morales (2007).
%\footnote{We use the ``case~1" \lx\ values from \citet{lopez07}. The results
%do not change significantly if we adopt the ``case~2" or ``case~3" \lx\ values instead.}
The sample includes 11 individual stars in 7 EB systems spanning the range 
\teff=3125--5300~K and $M$=0.21--0.96~\msun. 
We begin with the correlation of $\Delta R$ vs.\ \lx/\lbol\
already demonstrated in that work, which we rederived using the fundamental 
stellar data compiled in L\'opez-Morales (2007) and the same 3~Gyr isochrone 
of Baraffe et al.\ (1998)
as above. L\'opez-Morales (2007) did not discuss the complementary correlation with
$\Delta$\teff, but this information is also contained in the EB data,
so we also derive the relationship $\Delta$\teff\ vs.\ \lx/\lbol. 
%again using the data compiled in \citet{lopez07} and the 3~Gyr isochrone of \citet{baraffe98}.
Next we use the empirical \lx/\lbol\ vs.\ \lha/\lbol\ relationships of
Scholz et al.\ (2007) and Delfosse et al.\ (1998) convert the above from relationships
on \lx/\lbol\ into relationships on \lha/\lbol. The resulting relationships between
$\Delta$\teff, $\Delta R$, and $\log$~\lha/\lbol\ (Fig.~\ref{fig:finalrelations})
are statistically significant with $>$95\% confidence.

\section{Empirical Relations Linking Radius Inflation and \teff\ Suppression
to \ha\ Emission\label{2m0535}}

Combining the results from the field M dwarfs and EBs, we obtain the following
final relations (in percent units): 
$$\frac{\Delta T_{\rm eff}}{T_{\rm eff}} = 
(-4.71\pm 2.33) \times (\log L_{{\rm H}\alpha} / L_{\rm bol} +4) + (-4.4\pm 0.6)$$ 
$$\frac{\Delta R}{R} = 
(15.37\pm 2.91) \times (\log L_{{\rm H}\alpha} / L_{\rm bol} +4) + (7.1\pm 0.6).$$

%\section{Application to \2m \label{2m0535}}

As an example of how the {\it apparent} mass of an object can be altered by activity,
we calculate how the observed H-R diagram positions of the two BDs in \2m\ are altered
by the above relations. 
In effect, we are seeing how 
the \2m\ system would appear in the H-R diagram were the system completely inactive.
We use the \lha/\lbol\ measurements of Reiners et al.\ (2007) shown 
in Fig.~\ref{fig:2m0535}.
%who found $\log$~\lha/\lbol\ = $-3.47$ for the active primary and 
%$\log$~\lha/\lbol\ $< -4.30$ (upper limit) for the inactive secondary.
%From Eqs.~\ref{eq:Trelation} and \ref{eq:Rrelation},
We find that the primary BD has been displaced by 
%%\sout{the resulting $\Delta$\teff\ for the primary is} 
$\Delta$\teff\ = $-6.9\pm1.4$\% and $\Delta R = 15.2\pm1.7$\%,
and the secondary BD by at most $\Delta$\teff $=-3.0\pm0.9$\% and 
$\Delta R < 2.5\pm1.1$\%
(implying nearly constant \lbol\ for both). 
%(again implying a nearly constant \lbol).
%For both BDs the $\Delta$\teff\ and $\Delta R$ 
%offsets are such that \lbol\ is preserved to within 1--3\%. 
Shifting the position of the primary BD accordingly (Fig.~\ref{fig:2m0535})
shows that, if it were inactive, it would be in
excellent agreement with the theoretically expected position for its known mass.
%The observed \lha/\lbol\ of the active primary is near the ``saturation" limit and therefore
%near the upper limit of observed activity levels in young low-mass objects, for which the
%\ha\ emission is found to be mostly non-variable (Bell et al.\ 2012). Thus we do not
%expect a large contribution to the uncertainty of the active primary in Fig.~\ref{fig:2m0535}
%due to \ha\ variability. 
%We do not shift the position of the secondary as its \lha\ is an upper limit only. 

\section{Impact on Inferred Initial Mass Functions}

The vast majority of masses for stars and BDs in young clusters can only
be determined by comparison with theoretical evolutionary tracks in H-R
diagrams, using either \lbol\ or \teff\ or both (e.g., Scholz et al.\
2012).  Therefore, if there is indeed a relation between magnetic activity
and $R$ inflation / \teff\ suppression, this will also affect estimates of
stellar and substellar masses derived from \teff, especially at young ages
when activity levels are high. Here we examine the two ways of
estimating masses---\teff\ and \lbol---that are commonly used in the literature 
and investigate the impact of magnetic activity on the derived masses.

First we use \teff\ to estimate masses from model isochrones at 1~Myr.
At each model mass, we consider a range of activity levels and apply
offsets to the model \teff\ based on our empirical relations above.
We then use these suppressed \teff\ values to estimate the masses that
would be inferred from the isochrone.  As expected, this procedure leads
to a systematic {\it underestimation} of the masses. At high levels of
activity, the effect can be substantial. For $\log$~\lha/\lbol\ = $-3.3$,
which corresponds to the saturation limit in young low-mass stars and in
young associations (Scholz et al.\ 2007), the mass estimates are a factor
of $\sim 2$ lower than for objects with low levels of magnetic activity
($\log$~\lha/\lbol\ $<-4.5$).
Note that because the mass--\teff\ relationship is less steep at older ages, 
the underestimation of mass is less severe at older ages ($>$100~Myr).

In the second test case, we use \lbol\ derived from $K$-band absolute
magnitudes to derive masses. In this case the influence of magnetic activity
is introduced by the \teff\ dependence of the bolometric correction,
since \lbol\ is (as discussed above) practically not affected.  We compute
spectral types from the suppressed values of \teff, and then compute $K$-band
bolometric corrections from the spectral types, and we combine these with
the model $K$-band absolute magnitudes to find \lbol. Finally, we estimated
masses from the model isochrone and \lbol.  This second method still leads
to an underestimate of the masses, but the effect is much smaller than
when directly estimating masses from \teff.  The change of the bolometric
corrections with increasing magnetic activity is quite small (at most 10\%),
resulting in relatively minor changes of a few percent in the mass estimate.

An important application of \teff-based mass estimates is the determination
of initial mass functions (IMFs) for young clusters. M-dwarfs 
with ages from 10 to 100\,Myr exhibit \ha\ emission in the range
$\log$~\lha/\lbol\ = $-4.2$ to $-3.3$ (Scholz et al.\ 2007). 
%For BDs at the same ages the available data are sparse but indicate an upper limit
%around $-3.7$ \citep{barrado03}. 
Thus, our analysis
implies that objects inferred to be BDs from \teff-based (or spectral type
based) mass estimates could actually be low-mass stars since masses will
be underestimated by up to a factor of 2.

We further evaluated the impact of this effect on measurements of the slope of the IMF, 
$\alpha$ (in $dN/dM \propto M^{-\alpha}$).
For this purpose, we assume a measured slope of $\alpha = 0.6$
%which is consistent with a number of studies in very young clusters
(see Scholz et al.\ 2012), 
calculate the IMF based on that slope for $M<0.6\,M_{\odot}$, 
correct the masses for a given level of \ha\ emission using our relations above, 
and re-determine the slope $\alpha$. 
%If magnetic 
%activity and thus the level of \ha\ emission is constant across the low-mass regime, 
%$\alpha$ is practically unchanged, because all 
%masses will be underestimated by about the same factor. 
%Based on the available measurements, it seems realistic to assume 
%that BDs have on average a lower level of activity than low-mass stars 
If we start with $\log$~\lha/\lbol\ = $-4.0$ for BDs and $-3.5$ for low-mass stars
(Barrado et al.\ 2003),
the masses have to be corrected by factors of 1.3--1.7 for
BDs, and by 2.2--2.5 for low-mass stars. As a result, the slope changes 
from $\alpha = 0.6$ to $\alpha = 0.5$. 
%A more extreme mass dependence
%of the activity level will enhance this effect. 
In general, we expect that $\alpha$ will be underestimated by $\ga 0.1$, if the
masses are estimated from \teff\ and activity is not taken into account. 
In addition, the peak mass of the IMF 
would be underestimated by up to a factor of 2. 
%For a more quantitative assessment of these effects a larger sample 
%of \ha\ measurements for young stars and BDs is needed.

\section{Discussion\label{summary}}

It remains unclear what is the underlying physical mechanism driving the
correlations between $\Delta$\teff, $\Delta R$, and \lha. 
Chabrier et al.\ (2007) and MacDonald \& Mullan (2009) have suggested that a
sufficiently strong field could suppress convection, inhibit heat transfer,
and thus inflate (and cool) the stellar surface. Since such a field would
also likely result in chromospheric activity, one might therefore expect
the correlations with \ha\ emission that we have derived.

Browning (2008) has shown in global numerical models that fully
convective stars can host large kG strength fields.  However, such fields
alone appear to be too weak to produce the radius inflation and \teff\
suppression observed in \2m\ (MacDonald \& Mullan 2009). 
Chabrier et al.\ (2007) also
suggest that fully convective objects should be less affected by the
same convective inefficiencies invoked to explain radius discrepancies at
higher masses.  
It is possible that a combination of rotation and magnetic
activity contribute to both inflation/suppression and \ha\ emission in such
a way as to produce our empirical relation without a causal correspondence
between \ha\ and the magnetic field.

An alternative explanation for \teff\ suppression is a spot covered surface. 
%For the case of \2m, Mohanty et al.\ (2010) have argued that a model
%for the active primary with 70\% (axisymmetric) spot coverage could explain
%the peculiar mass-temperature relationship in that system. 
Since spot coverage is also controlled by magnetic fields, a correlation with \ha\
might still exist. 
%It remains unclear how to interpret spot coverage of
%greater than 50\%; perhaps the analogy with solar type spots breaks down
%in such extreme systems.  
However, Mohanty et al.\ (2010) and Mohanty \& Stassun (2012) have now shown from 
a spectral fine-analysis of the \2m\ system, observed at high resolving power
during both primary and secondary eclipses, that such a large-spot scenario
is strongly disfavored as an explanation for the \teff\ suppression of the
primary BD in \2m. Thus, while the \teff\ suppression mechanism produces a
clear correlation with chromospheric \ha\ activity as we have shown here,
it evidently does not in all cases effect this correlation directly through
surface spots.

\section{Conclusions\label{conclude}}

We have shown that there exists a correlation between the strength of \ha\
emission in active M-dwarfs, and the degree to which their \teff\ are
suppressed and radii inflated compared with inactive objects and theoretical
evolutionary models.  
Our relationships above should prove directly
useful for a variety of applications in which accurate estimates of masses and radii
are needed for low-mass stars and BDs, and should assist in correcting inferred
initial mass functions at young ages.

While promising, the correlations we have derived
contain significant scatter, and they are currently limited by the lack
of a single sample of stars with both \ha\ and direct radius measurements.
We therefore encourage researchers to publish \ha\ measurements,
as this is usually available from the spectra used to determine EB
radial velocities.

Finally, the relations we have determined already indicate quite clearly that
the radius inflation and temperature suppression mechanism operates in
such a way that the temperature suppression and radius inflation almost
exactly cancel in terms of their effect on the bolometric luminosity.
Moreover, the relations between activity, \teff\ suppression, and radius
inflation do not appear to manifest any obvious discontinuity across the
fully convective transition (see also Stassun et al.\ 2010).  These are
important, fundamental clues to the physical nature of these effects,
and should help to constrain theoretical models that are being developed
to explain these phenomena (e.g., Chabrier et al.\ 2007; MacDonald \& Mullan 2009).

%\acknowledgements
%We appreciate the cooperation with the authors of AN. Based on
%photographic data obtained using the UK Schmidt Telescope, etc..

% Use this code if you wish to generate your bibliography with BibTeX;
% please replace first the string "an-demo" below with the name(s) of
% the BibTeX data base(s) you want to use.
% The resulting bibliography-output (the contents of the .bbl file)
% must be pasted into this file before submission.
%
% \bibliographystyle{an}
% \bibliography{an-demo}
%
% Replace the following example bibliography with your references
% before submission:

\end{document}